\documentclass[conference]{IEEEtran}
\IEEEoverridecommandlockouts
\usepackage{cite}
\usepackage{amsmath,amssymb,amsfonts}
\usepackage{algorithmic}
\usepackage{graphicx}
\usepackage{textcomp}
\usepackage{xcolor}
\def\BibTeX{{\rm B\kern-.05em{\sc i\kern-.025em b}\kern-.08em
    T\kern-.1667em\lower.7ex\hbox{E}\kern-.125emX}}
\begin{document}

\title{Non-verbal Facial Action Units-based Automatic Depression Classification\\

\thanks{Identify applicable funding agency here. If none, delete this.}
}

\author{\IEEEauthorblockN{ Chuang Yu}
\IEEEauthorblockA{\textit{Shenzhen Institute of Advanced Technology }\\
Shenzhen, China \\
chuang.yu@siat.ac.cn 
}
}

\maketitle

\begin{abstract}
Depression is a common mental disorder that causes people to experience depressed mood, loss of interest or pleasure, feelings of guilt or low self-worth. Traditional clinical depression diagnosis methods are subjective and time consuming. Since depression can be reflected by human facial expressions, We propose a non-verbal facial behavior-based automatic depression classification approach. In this paper, both short-term behavior-based and clip-based depression classification are constructed. The final clip-level decision of short-term behavior-based depression detection is yielded by averaging the predictions of all short-term behaviors while we modelling behaviors contained in all frames based on two Gaussian Mixture Models. To evaluate the proposed approaches, we select a gender balanced subset from AVEC 2019 depression corpus containing 30 participants. The experimental results show that our method achieved more than 75\% depression classification accuracy, where both GMM-based clip-level depression modelling and rank pooling-based short-term depression behavior modelling achieved at least 70\% classification accuracy. The result indicates that our approach can leverage complementary information from both systems to achieve promising depression predictions from facial behaviors.
\end{abstract}



\section{Introduction}

\noindent Depression is a common mental disorder that causes people to feelings of guilt or low self-worth, disturbed sleep or appetite, low energy, and poor concentration. Accurate diagnosis of MDD requires intensive training and experience. Thus the growing global burden of depression suggests that an automatic means to monitor depression severity would be a beneficial tool for patients, clinicians, and healthcare providers \cite{sturim2011automatic}. Therefore, it is important to explore the automatic way for depression assessment, which would allow patients to be assessed by AI systems that do not need long-term training. Recent literature \cite{song2020spectral,de2021mdn,song2018human,uddin2020depression,yang2018integrating,eye1stolicyn2020prediction,jaiswal2019automatic} suggested that human non-verbal behaviors such as facial actions, speech rate, etc, are informative of depression cues. Because of this, many non-verbal behaviors based depression detection systems were developed.

While Cohn et al. \cite{cohn2009detecting} is a pioneer work in the automatic depression recognition area, it fused both visual and audio modality together in an attempt to incorporate behavioral observations, from which are strongly related to psychological disorders. Recent studies can be divided into two categories based on their spatio-temporal modelling solution: 1. static image-based depression prediction; 2. image sequence-based depression prediction. The static image-based approaches \cite{mcintyre2009approach,cootes2001active,jaiswal2019automatic,shin2016deep,zhou2018visually,he2021automatic} attempts to predict depression status from a single face image without considering spatio-temporal facial behaviors. For the task that videos are provided, these approaches usually first make a prediction for each frame and then combine predictions of all frames to make final video-level prediction. While such static image-based approaches cannot represent facial dynamics that have been claimed to be crucial for depression diagnose \cite{gehricke2000reduced,carney1981facial}, they are not reliable for depression recognition. In other words, depressed patients may not show depressive behaviors in all frames while non-depressed individuals sometimes display depressive facial expressions.

To use these important facial dynamics, some image sequence-based approaches propose to divide an video into several short segments with a fixed number of frames, and then extract a feature for each segment. In particular, these methods usually specifically model the spatio-temporal patterns of each segment using Space-Time Interest Points \cite{joshi2013can}, Feature Dynamic History Histogram \cite{jan2017artificial}, 3D CNN\cite{de2021mdn}, etc., which allow the short-term facial dynamics to be used for depression recognition. However, due to the high dimensionality of the video, it is not easy to directly feed an entire video into a depression model. Therefore, the aforementioned approaches still ignore the long-term facial behavioral dynamics of the video. While depression is a long-term mental status, the long-term facial behavioral dynamics would also contribute significant information for its recognition \cite{song2018human,song2020spectral,xu2021two}.

Since facial action units (AUs) can be accurately and automatically detected from the face sequence, this paper presents a novel approach that automatically learns dynamic feature representations from mutiple AU time-series, aiming to predicting the depression status of individuals. Specifically, we first divide the whole time-series of each individual into multiple short segments, and then propose to capture the short-term temporal ordering of a facial AU behaviors by training a linear ranking machine \cite{fernando2016rank} on each short segment. Thus, the parameters of the linear ranking function encode temporal evolution of facial behaviors contained in each segment, which can be treated as a short-term facial dynamic representation for the corresponding segment. Meanwhile, we also learn two Gaussian Mixture Models (GMM) to represent the whole distribution of depressed and non-depressed individuals' AU time-series, which aims to predict individuals' depression status based on the clip-level information. The main contribution of this work is that we provide a new solution to jointly that can encode both short-term and clip-level facial AU temporal information for depression classification. To the best of our knowledge, this is the first approach that applies rank pooling and GMM to encode depression features from facial AUs.

\section{Related Work}

\subsection{Face-based depression recognition}

\noindent  Since facial behaviors are informative for reflecting depression symptoms \cite{de2020deep,uddin2020depression,zhou2018visually,song2020spectral,al2018video,10.1145/2512530.2512533,9053207,201410.1145/2661806.2661807,gratch2014distress}, current depression recognition studies can be divided into two categories based on their spatio-temporal modelling solution: 1. facial appearance (static face)-based depression prediction; 2. facial dynamics-based depression prediction.  

While the static face-based approaches only considered static facial displays and ignores the temporal correlations between frames, Joshi et al. \cite{joshi2013can} is an early approach that extract spatio-temporal pattern for depression recognition, which analyzes the changes of the head and facial landmarks across both spatial and temporal dimensions. Then, a histogram is constructed based on the found Space-Time Interest Points (STIP). The experiments on a clinical database show that the changes in STIP of normal people and depression patients are significantly different. Zhu et al.\cite{zhu2017automated} represents temporal information by extracting Dynamics changes of adjacent frames. They express facial features by fine-tuning the pre-trained DCNN. Then, because the facial feature changes between adjacent frames are very small, an additional DCNN captures the movement of the face by calculating the optical flow between several consecutive frames. Jan et al. proposes a Feature Dynamic History Histogram \cite{jan2017artificial}, which firstly uses a pre-trained AlexNet \cite{krizhevsky2012imagenet} to extract the static facial feature for each frame. Then, the grey scale value changes of all pixels are represented as Feature Dynamic History Histogram, encoding facial temporal information. The static and dynamic visual features as well as audio feature are then combined  through a linear regression model to predict depression. However, exploiting spatial and temporal information separately can deteriorate the modeling of spatio-temporal relationships\cite{de2019combining}. To summarize multi-scale video-level facial dynamics, Song et al. \cite{song2020spectral,song2018human,xu2021two} propose to convert AUs time-series of the video to frequency domain, i.e, a spectral signal of the video. Then, a frequency alignment is proposed to convert the produced spectral signal to a fixed-size video-level spectral representation, where each retained frequency represent a unique video-level dynamic.

Spatio-temporal CNNs also have been widely used in recent years to jointly deep learn spatio-temporal information from the video. Jazaery et al. \cite{al2018video} proposed a two-stream C3D model to jointly extract the spatio-temporal features from both aligned and Non-aligned face clips. In order to conduct the depression prediction task with limited training depression videos, both C3D models was pre-trained on the Sports-1M dataset and then fine-tuned using UCF101 human action dataset \cite{tran2015learning}. Finally, the spatial and temporal features extracted by two C3D models are fused by an RNN model to provide final depression prediction. Melo et al. also used two-stream C3D model to extract the spatial-temporal facial features to predict depression \cite{de2019combining}. However, due to the limited training data, the employed 3D CNNs are always required to be pre-trained, which may suffer from overfitting. Their later work \cite{de2021mdn} also employs C3D-style network for depression recognition, where a Maximization-Differentiation module is proposed to enhance the depression-related feature learning process. Importantly, this work firstly down-samples the entire video into a small number of frames as the video representation and then feeds all of them to the network to recognize depression at the video-level.

\subsection{Rank Pooling}

\noindent Temporal information is crucial for continuous time-series data analysis, as the the target signal at a certain time usually depend on its previous states. The standard solution for modelling temporal dependencies is through sequential latent models such as Recurrent Neural Networks (RNN), or temporal convolution neural network \cite{lea2016temporal}. However, such models are hard to train as they are usually sensitive to hyper-parameters as well as suffering from overfitting. Moreover, these methods usually learn temporal information from extracted features rather than original data themselves, which may lead to the losses of key information. Instead, we draw our attention to a recent proposed novel rank pooling strategy~\cite{fernando2015modeling,fernando2016rank} that summarizes the temporal evolution of a time-series signal into a RankSVM~\cite{smola2004tutorial} kernel, which is learned to rank all frames based on their temporal orders. More importantly, this technique does not require human annotations for learning temporal representation of the time-series sequence, making it practically useful for real-world applications. While the rank pooling technique has been further extended and achieved state of the art performance for many computer vision problems such as action recognition \cite{bilen2017action}, facial expression analysis \cite{song2019dynamic,song2021selfsuper}, gesture recognition \cite{wang2016large,wang2017large}, personality recognition \cite{beyan2019personality,song2021self}, etc., its application to 1-D financial time-series data analysis haven't been fully explored. In this paper, we extend the rank pooling algorithm to stock prices forecasting task by constructing multi-scale dynamic representations.

\section{The proposed approach}

\begin{figure*}
\centering
\includegraphics[width=17cm]{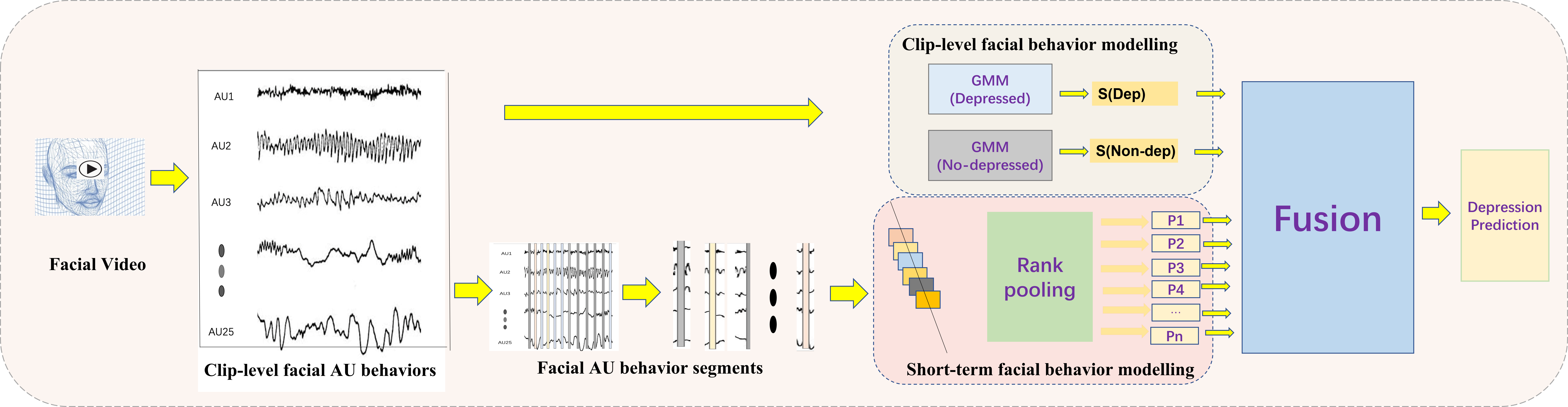} 
\caption{The pipeline of the proposed approach.}
\label{pipeline}
\end{figure*}

\noindent In this section, we propose a hybrid system that utilizes Gaussian Mixture Model (GMM) for clip-level facial behavior depression feature extraction, and Rank pooling algorithm for short-term facial behavior depression feature extraction, where 17 facial action units intensities are used as the frame-level facial descriptors. The whole pipeline of our approach is demonstrated in Fig. \ref{pipeline}.

\subsection{Gaussian Mixture Model-based clip-level facial representation learning}


\noindent Based on the Bayesian posterior probability framework, the GMM model represents the distribution of training facial AU time-series data by the sum of multiple Gaussian models, of which the parameters (mean and standard deviation of Gaussian models) can fit to the data using Expectation-Maximum (EM) algorithm. Here, we use the GMM model consists of $n$ Gaussian components ($n$ centers) to represent the data:
\begin{equation}
p(x) = \sum_{k=1}^n \pi_k N(x|\mu_k,\Sigma_k)
\end{equation}
where $x$ is the facial AU data; $\pi_k$ is the mixing coefficient; $N(x|\mu_k,\Sigma_k)$ is the Gaussian model. In the testing phase, the log of likelihood between facial AU time-series of the test data and the GMM model is given by
\begin{equation}
\text{ln}p(X|\pi,\mu,\Sigma) = \sum_{m=1}^M \text{ln} (\pi_k N(x|\mu_k,\Sigma_k))
\end{equation}
where $X$ is the test facial AU data, $\pi$, $\mu$ and $\Sigma$ are the parameters of GMM model. It applies not only mean values but also weights and standard deviation values. This property enables the GMM model to represent training data accurately. In this paper, we use two GMM model with 32 centers, representing depressed and non-depressed participants, respectively. As a result, the clip-level facial information of each participant can be compared with the trained depressed and non-depressed GMM models.




\subsection{Rank pooling-based short-term facial representation learning}

\noindent We then extend the rank-pooling algorithm to encode short-term facial behavior from the AU time-series using a time-window. We assume that each AU time-series of a given time window is represented by a vector $\textbf{x}$. Then, a clip can be represented a set of stacked AU segments  $X=\left[\mathrm{\textbf{x}}_{1}, \mathrm{\textbf{x}}_{2}, \ldots, \mathrm{\textbf{x}}_{N}\right]$.  A sequence of a given window of time can be represented at the time $t$ as $\mathrm{\textbf{x}}_{\mathrm{{t}}} \in \mathbb{R}^{17 \times D}$ (i.e., $\mathrm{\textbf{x}}_{\mathrm{{t}}}$ has $D$ frames, each frame is described by $17$ facial AUs). We use the notation $\mathrm{x}_{t: t+D}$ to denote a sub-sequence from time step $t$ to $t+D$. The goal is to extract facial dynamics $\mathcal{D}_t$ of each segment $\mathrm{\textbf{x}}_{t: t+D}$ for depression classification. To achieve this, we train an RankSVM kerenl $d_t$ that can rank all frames in the sequence according to their temporal orders by assigning ascending scores, which is defined as
\begin{equation}
\label{eq:score}
S(a) = \langle d_t, X_a \rangle
\end{equation}
which is conditioned on the pairwise constraints:
\begin{equation}
S(a) - S(b) > T \vert a > b
\end{equation}
where $X_a$ is the $a_{th}$ frame in the segment, and $\langle d_t, X_a \rangle$ denotes the dot product between the kernel and the facial attributes of the $a_{th}$ frame; $T$ is a threshold. As a result, $d_t$ would reflect facial dynamics of the whole segment, as it enforces for all frames in the sequence to be placed in the correct temporal order. As a result, we learn a set of short-term facial dynamic representations $d_1, d_2, \cdots, d_N \in \mathbb{R}^{1 \times 17}$ to describe all facial behavior segments of a clip. We train a three layer Multi-Layer Perceptron (MLP) for short-term facial behavior representation-based depression classification, where the dropout rate between each pair of layers is 0.5 and the ReLU activation function is also employed.

\subsection{Combining short-term and clip-level facial behavior representations for depression classification}

\noindent For a given clip that consists of $N$ short segments, $N$ short-term depression classification results will be obtained. We propose to combine them as well as the log likelihoods obtained between the clip-level facial AU time-series and two GMM models to make the final classification prediction as:
\begin{equation}
 P = \text{ln}p(X|\pi,\mu,\Sigma)_\text{Dep} - \text{ln}p(X|\pi,\mu,\Sigma)_\text{N-Dep} + \omega \sum_{i = 1}^{N} P(d_i)
\end{equation}
where $P(d_i) = 1$ if the segment is predicted as depressed and $P(d_i) = 0$ otherwise; $\omega$ is used for adjusting the weights of the predictions made by short-term and clip-level facial behaviors.

\section{Experiments results}

\subsection{Dataset}

\noindent In this paper, we build a subset from the AVEC 2019 depression dataset \cite{ringeval2019avec} to construct a balanced dataset, which contains 15 males and 15 females. Their average age is 28.5 years old. The depression state of each participant was categorized to be either depressed or Non-depressed based on the provided PHQ-8 score. All videos were collected based on interviews conducted by an animated virtual interviewer called Ellie. During the interview, the Ellie is either controlled by a human interviewer (wizard) in another room, or Ellie acts in a fully autonomous way using different automated perception and behaviour generation modules. In this paper, we used the 17 AU intensities provided in the dataset as frame-level human facial descriptors.

\subsection{Implementation details}

\noindent We conducted leave-one-out scheme for 30 samples. At each step, 29 clips, were used for training while another clip is for testing. Table \ref{tab1} shows the classification results for all participants obtained from specialists and two automatic methods. In addition, we also conducted three baselines for comparison as they either provide the code online or easy to reproduce. These include: 1. the Dilated Temporal Convolution Neural Network used by \cite{haque2018measuring}; 2. the Spectral Vector used by \cite{song2020spectral}; 3. the MLP used by \cite{jaiswal2019automatic}. The evaluation metrics is the binary classification accuracy.

\subsection{Results and discussion}

\noindent As shown in Table \ref{comparsion}, it is clear that our method achieved much higher result ($76.7\%$) than the chance level ($50\%$) result. It is also need to emphasize that our method achieved better classification result over all compared approaches. This result can be explained as our method utilized both clip-level and short-term behavior information without losing any details or long-term dependencies. In comparison, Dilated CNN-based model only used short-term information. Meanwhile, Spectral Vector-based system and MLP-based system only modelled clip-level behavior features.

\begin{table}
\begin{center}
\caption{The detailed depression classification results}\label{tab1}
\begin{tabular}{|c c |}
\hline
Method & Accuracy \\
\hline
Dilated CNN \cite{haque2018measuring} &  70.0 \% \\
Spectral Vector \cite{song2020spectral} & 73.3\%  \\
MLP \cite{jaiswal2019automatic} & 60.0 \%    \\
Ours & 76.7\%  \\
\hline
\end{tabular}
\end{center}
\label{comparsion}
\end{table}

\begin{figure}
\centering
\includegraphics[width=8cm]{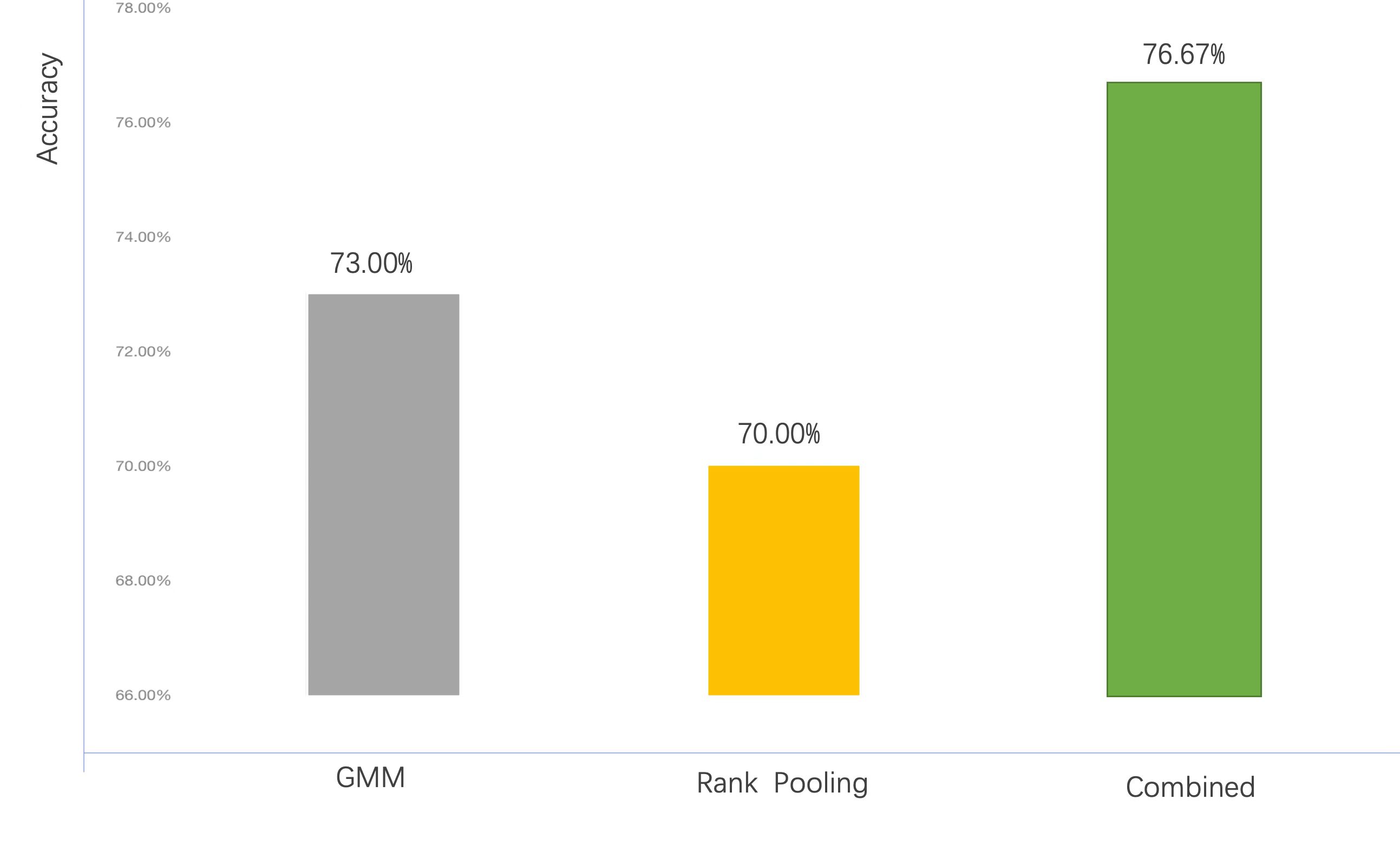} 
\caption{Comparison between three system settings.}
\label{Result}
\end{figure}

As shown in Table \ref{details} and Fig. \ref{Result}, we can also observed that combining our GMM-based clip-level depression prediction with the rank pooling-based short-term depression predictions can generate better results than using either of them only, i.e., the classification accuracy of the combined system (76.7\%) outperforms the GMM-based system (73\%) and rank pooling-based system (70\%). This show that both clip-level and long-term facial behaviors contain unique depression-related cues. More specifically, as we can see from Table \ref{details}, when one system is incorrectly classified the depression status of the clip while the other generated correct prediction, our combined system usually output the correct prediction (such as ID 10 and ID 17). Also, both GMM-based and rank pooling-based systems achieved over $50\%$ classification accuracy (the chance level accuracy), i.e., both of them achieved at least than 70\% classification accuracy. As a result, it can be concluded both of them are able to extract depression-related cues from the raw AU data.

\begin{table*}
\begin{center}
\caption{The detailed depression classification results}\label{tab1}
\begin{tabular}{|c c c c c|}
\hline
Participant ID. & Label &  Result(GMM) & Result(Rank Pooling) & Result(Combined) \\
\hline
1& Depressed & Depressed & Depressed & Depressed  \\
2& Depressed & Non-depressed & Depressed & Depressed  \\
3& Non-depressed & Non-depressed & Non-depressed & Non-depressed  \\
4& Depressed & Depressed & Depressed & Depressed \\
5& Non-depressed & Non-depressed & Non-depressed & Non-depressed \\
6& Non-depressed & Non-depressed & Non-depressed & Non-depressed \\
7& Non-depressed & Non-depressed & Depressed & Non-depressed \\
8& Non-depressed & Depressed & Depressed & Depressed \\
9& Non-depressed & Non-depressed & Non-depressed & Non-depressed \\
10& Depressed & Non-depressed & Depressed & Depressed \\
11& Non-depressed & Depressed & Depressed & Depressed \\
12& Non-depressed & Non-depressed & Non-depressed & Non-depressed \\
13& Depressed & Depressed & Non-Depressed & Non-depressed  \\
14& Depressed & Non-depressed & Non-depressed & Non-depressed \\
15& Depressed & Depressed & Depressed & Depressed \\
16& Non-depressed & Non-depressed & Non-depressed & Non-depressed \\
17& Depressed & Depressed & Non-depressed & Depressed \\
18& Non-depressed & Non-depressed & Non-depressed & Non-depressed\\
19& Non-depressed & Non-depressed & Depressed & Depressed \\
20& Depressed &  Depressed & Depressed & Depressed \\
21& Depressed & Depressed & Depressed & Depressed \\
22& Non-depressed & Depressed & Non-depressed & Non-depressed \\
23& Depressed & Depressed & Non-Depressed & Depressed  \\
24& Depressed & Non-depressed & Depressed & Non-depressed \\
25& Non-depressed & Non-depressed & Non-depressed & Non-depressed \\
26& Depressed & Depressed & Depressed & Depressed \\
27& Depressed & Depressed & Depressed & Depressed \\
28& Non-depressed & Depressed & Depressed & Depressed\\
29& Non-depressed & Non-depressed & Non-depressed & Non-depressed \\
30& Depressed &  Depressed & Depressed & Depressed \\
\hline
\end{tabular}
\end{center}
\label{details}
\end{table*}

\section{Conclusion}

\noindent This paper propose a combined depression classification system that extends GMM to model clip-level facial behavior features and the rank pooling to model short-term facial behavior features. The results show that our approach can effectively extract and combine depression-related clip-level and short-term facial behavior features, and achieved superior depression classification result. This means it is worth to further develop them as the external assistant system for our clinical depression assessment.


\bibliographystyle{IEEEtran}
\bibliography{ref}

\end{document}